\documentstyle[twocolumn,aps,epsf]{revtex}

\begin{document}
\draft
\twocolumn[\hsize\textwidth\columnwidth\hsize\csname@twocolumnfalse\endcsname
\title{A renormalisation-group treatment of two-body scattering}
\author{Michael C. Birse,  Judith A. McGovern and Keith G. Richardson}
\address{Theoretical Physics Group, Department of Physics and Astronomy\\
University of Manchester, Manchester, M13 9PL, UK\\}
\maketitle
\begin{abstract}
Nonrelativistic two-body scattering by a short-ranged potential is studied
using the renormalisation group. Two fixed points are identified: a
trivial one and one describing systems with a bound state at zero energy. The
eigenvalues of the linearised renormalisation group are used to assign a
systematic power-counting to terms in the potential near each of these fixed
points. The expansion around the nontrivial fixed point is shown to be 
equivalent to the effective-range expansion.
\end{abstract}
]

Recently there has been much interest in applying the techniques of effective
field theory (EFT) to the scattering of massive particles interacting via
short-ranged forces. This has been spurred by Weinberg's power-counting rules
for the low-momentum expansion of the nucleon-nucleon potential\cite{wein2},
which raised the possibility of applying the techniques of chiral perturbation
theory to nuclear physics\cite{orvk}. These would provide a systematic method
for expanding few-nucleon bound-state properties and scattering observables in
powers of nucleon momenta and the pion mass.

By focussing on the potential, one avoids contributions where the two
intermediate nucleons are almost on-shell, giving small denominators. However,
this is the physics responsible for nuclear binding, and so to describe nuclei
with an EFT it is not enough to write down a potential; one needs to solve the
corresponding Schr\"odinger or Lippmann-Schwinger equation. At this point one
encounters a problem. The EFT is based on a Lagrangian with local couplings
between the particles and these include contact interactions between the
nucleons. Such interactions correspond to $\delta$-function potentials, and
the resulting scattering equations only make sense after a further
regularisation and renormalisation.

A variety of approaches has been explored for renormalising two-body
scattering by such potentials
\cite{afg,ksw,md1,md2,lep,vk1,rbm,vk2,bvk,pkmr,ksw2,geg,vk3}. (For reviews
of the various approaches and further references, see Ref.~\cite{npeft}.) 
These have shown that it is difficult to set up a useful and systematic EFT for 
two-body scattering when the scattering length is unnaturally large, as indeed 
noted by Weinberg\cite{wein2}. More recently an alternative to Weinberg's power 
counting has been suggested by Kaplan, Savage and Wise (KSW)\cite{ksw2}, based 
on dimensional regularisation with a ``power divergence subtraction" scheme 
(PDS). The same counting has also been proposed by van Kolck\cite{vk2,vk3} 
within the framework of a general subtractive renomalisation scheme, using a 
momentum cut-off set at a large scale, typical of the underlying physics 
responsible for the short-distance interactions.

To examine some of the questions raised by these approaches, we have studied
nonrelativistic two-body scattering from the viewpoint of Wilson's continuous
(or the ``exact") renormalisation group (RG)\cite{wrg}. In this approach, one
imposes a cut-off on the momenta of virtual states at some scale $\Lambda$ and
demands that physical quantities be independent of $\Lambda$. A rescaled
Hamiltonian is introduced in which all dimensioned variables are expressed in
units of $\Lambda$. The RG then describes the variation with $\Lambda$ of this
rescaled Hamiltonian.

Of particular interest in the present context is the behaviour as the cut-off
is taken to zero. In this limit, more and more physics is integrated out and
its effects are included implicitly in the effective Hamiltonian (in the
present case, two-body potential). Eventually only $\Lambda$ is left to set an
energy scale and as a result the rescaled potential becomes independent of
$\Lambda$. This is an infra-red fixed point of the RG. In the case of two-body
scattering we find two types of fixed point, which correspond to two possible
scale-independent forms for the zero-energy scattering amplitude, expressed in
units of $\Lambda$. One is a trivial fixed point with a vanishing scattering
amplitude. The others are a set of fixed points which give bound states at
exactly zero energy.

Low-energy scattering (in physical units) can then be described in terms of
perturbations of the potential around one of the fixed points. These
perturbations can be expanded in terms of eigenfunctions of the linearised RG
equations, which scale with definite powers of $\Lambda$. The counting of
powers of $\Lambda$ in these eigenfunctions provides a systematic way to
organise the terms in the potential according to how rapidly they head
towards, or away from, the fixed point. The eigenvalues also show whether the
fixed point is stable or not.

For two-body scattering we find that the trivial fixed point is stable and
that perturbations around it have scaling dimensions given by Weinberg's power
counting rules\cite{wein2}. In contrast the fixed-points with zero-energy
bound states are unstable. We find that each (energy-dependent) perturbation 
around the least unstable of these points corresponds to a term in the
effective-range expansion\cite{ere}. The power counting for these
perturbations agrees with that suggested by KSW\cite{ksw2} and van
Kolck\cite{vk3}. For a purely short-range potential, there is thus an
equivalence between the effective field theory and the effective-range
expansion\cite{ere,bl}.

We consider here $s$-wave two-body scattering by a potential that consists of
contact interactions only. In general this potential depends on energy as well
as on the initial and final relative momenta. To second order in the momentum
expansion it has the form
\begin{equation} \label{eq:pot1}
V(k',k,p)=C_{00}+C_{20}(k^2+k'^2)+C_{02}\,p^2\cdots,
\end{equation}
where, as throughout this paper, we use $k$ and $k'$ to denote relative momenta
and the energy-dependence is expressed in terms of $p=\sqrt{ME}$, the on-shell
momentum corresponding to the energy $E$. Unlike previous applications of RG 
ideas to two-body scattering\cite{wein2,afg,ksw2}, we do not restrict our 
potential to only leading or next-to-leading terms in the momentum expansion.

In treating the scattering non-perturbatively, it is convenient to work with
the reactance matrix, $K$. The off-shell $K$-matrix satisfies a
Lippmann-Schwinger (LS) equation that is very similar to that for the
scattering matrix $T$, except that the Green's function satisfies
standing-wave boundary conditions:
\begin{eqnarray} \label{eq:lse}
K(k',k,p)&=&V(k',k,p)\\
&&+\frac{M}{2\pi^2}{\cal{P}}
\int_{0}^{\Lambda}q^2dq\,\frac{V(k',q,p)K(q,k,p)}
{{p^2}-{q^2}},\nonumber
\end{eqnarray}
where ${\cal P}$ denotes the principal value of the integral. We have chosen to
regulate the integral here by imposing a sharp cut-off at $q=\Lambda$. The
inverse of the on-shell $K$-matrix differs from that of the on-shell
$T$-matrix by a term $iM p/4\pi$, which ensures that $T$ is unitary if $K$ is
Hermitian. This allows the effective-range expansion\cite{ere} to be written
as an expansion of $1/K$:
\begin{equation} \label{eq:ere}
\frac{1}{K(p,p,p)}
=-\frac{M}{4\pi}\left(-\frac{1}{a}+\frac{1}{2}r_{e}p^2
+\cdots\right),
\end{equation}
where $a$ is the scattering length and $r_e$ is the effective range.

The RG equation for the potential is obtained by making $V$ dependent on
$\Lambda$ and demanding that the off-shell $K$-matrix be independent of
$\Lambda$. This is obviously sufficient to ensure that all scattering 
observables do not depend on $\Lambda$. By differentiating the 
LS equation (\ref{eq:lse}) with respect to $\Lambda$,
setting $\partial K/\partial\Lambda=0$ and then operating
from the right with $(1+G_0K)^{-1}$ (where $G_0$ is the free Green's function
in Eq.~(\ref{eq:lse})), one obtains the equation
\begin{equation}\label{eq:rge}
{\partial V\over\partial\Lambda} 
={M\over2\pi^2}V(k',\Lambda,p,\Lambda){\Lambda^2\over\Lambda^2-p^2}
V(\Lambda,k,p,\Lambda).
\end{equation}
If this effective potential is to describe scattering by short-ranged 
interactions, it should be an analytic function of $k^2$ and $k'^2$ for small
$k$ and $k'$.\footnote{Rotational invariance also allows dependence on ${\bf k}
\cdot{\bf k}'$, but this is needed only for partial waves with nonzero angular
momentum.} Also, if the energy lies below all thresholds for production of
other particles then the potential should be an analytic function of the
energy, $E=p^2/M$. Under these restrictions, the boundary conditions that we
impose on $V$ are that it should have an expansion in non-negative, integer 
powers of $k^2$, $k^{\prime 2}$ and $p^2$.

We now introduce dimensionless momentum variables, $\hat k=k/\Lambda$ etc.,
and a define a rescaled potential,
\begin{equation}
\hat V(\hat k',\hat k,\hat p,\Lambda)={M\Lambda\over 2\pi^2}
V(\Lambda\hat k',\Lambda\hat k,\Lambda\hat p,\Lambda).
\end{equation}
In terms of these quantities, the RG equation takes the form
\begin{eqnarray}\label{eq:scrge}
\Lambda{\partial\hat V\over\partial\Lambda}&=&
\hat k'{\partial\hat V\over\partial\hat k'}
+\hat k{\partial\hat V\over\partial\hat k}
+\hat p{\partial\hat V\over\partial\hat p}
+\hat V\nonumber\\
&&\qquad+\hat V(\hat k',1,\hat p,\Lambda){1\over 1-\hat p^2}
\hat V(1,\hat k,\hat p,\Lambda).
\end{eqnarray}

In what follows, the idea of a fixed point of the RG will be crucial. As the
cut-off $\Lambda$ is taken to zero, more and more physics is integrated out
until $\Lambda$ itself is the only remaining scale. In the limit $\Lambda
\rightarrow 0$, the rescaled potential, being dimensionless, must become
independent of $\Lambda$. This means that as $\Lambda$ varies the rescaled
potential must flow towards an infra-red fixed point of the RG 
equation (\ref{eq:scrge}). The fixed points are described by solutions
of this equation that satisfy
\begin{equation}
{\partial\hat V\over\partial\Lambda}=0.
\end{equation}
An obvious example is the trivial fixed point, 
\begin{equation}
\hat V(\hat k',\hat k,\hat p,\Lambda)=0.
\end{equation} 
The $K$-matrix for this potential is also zero, corresponding to no scattering.

If, for a particular system, we find that the rescaled potential tends towards
this fixed point as we lower the cut-off towards zero, then we can describe the
low-energy behaviour in terms of the perturbations that scale with definite
powers of $\Lambda$. To find these we linearise the RG equation
(\ref{eq:scrge}) about the trivial fixed point and look for solutions of the
form 
\begin{equation}
\hat V(\hat k',\hat k,\hat p,\Lambda)=C\Lambda^\nu \phi(\hat k',\hat
k,\hat p),
\end{equation}
where the functions $\phi$ satisfy the eigenvalue equation
\begin{equation}\label{eq:linrge.tr}
\hat k'{\partial\phi\over\partial\hat k'}
+\hat k{\partial\phi\over\partial\hat k}
+\hat p{\partial\phi\over\partial\hat p}
+\phi=\nu\phi.
\end{equation}
It is not hard to see that the solutions to this that are well-behaved as 
the momenta and energy tend to zero are 
\begin{equation}
\phi(\hat k',\hat k,\hat p)=\hat k^{\prime l}\hat k^m \hat p^n,
\end{equation}
with RG eigenvalues $\nu=l+m+n+1$, where $l$, $m$ and $n$ are non-negative even
integers. The RG eigenvalues are all positive and so the fixed point is a
stable one: starting from any potential in the vicinity of $\hat V=0$ the
RG flow will take the potential to the fixed point as $\Lambda\rightarrow 0$.

The momentum expansion of the full potential near the trivial fixed point 
can be written 
\begin{equation}\label{eq:potexp.tr}
\hat V(\hat k',\hat k,\hat p,\Lambda)=\sum_{l,n,m}\widehat C_{lmn}\left(
{\Lambda\over\Lambda_0}\right)^\nu \hat k^{\prime l}\hat k^m \hat p^n.
\end{equation}
For a Hermitian potential one can take real coefficients with
$C_{lmn}=C_{mln}$.\footnote{Terms with forms like $i(k^2-k^{\prime 2})$ need
not be included since they can be shown to vanish when integrated by parts in
coordinate space.} We have chosen to write the coefficients in this expansion
in a dimensionless form by taking out  a factor of $\Lambda_0^{-\nu}$, where
$\Lambda_0$ is some scale associated with the underlying physics that
determines where our momentum expansion of the potential breaks down. This can
be seen more clearly from the form of the corresponding unscaled potential,
\begin{equation}\label{eq:potexp.utr}
V(k',k,p,\Lambda)={2\pi^2\over M\Lambda_0}\sum_{l,n,m}\widehat C_{lmn}
{k^{\prime l}k^m p^n\over\Lambda_0^{l+m+n}}.
\end{equation}
In a ``natural" theory it is possible to choose the scale $\Lambda_0$ in such a
way that all of the dimensionless coefficients are of order unity. The power
counting in this expansion of the potential (\ref{eq:potexp.tr}) is just the
one proposed by Weinberg\cite{wein2} if we assign an order $d=\nu-1$ to each
term.

The behaviour near the trivial fixed point gives weak scattering at low
energies which can be treated perturbatively. It can be used to describe
systems where the scattering length is small. For such systems Weinberg's
power counting has already been shown to provide a systematic treatment in the
context of both dimensional regularisation with minimal subtraction\cite{ksw}
and cut-off approaches\cite{md2} where the cut-off is chosen to be well below
$1/a$ and $1/r_e$. In these cases the $K$-matrix is given by the first Born 
approximation and so is equal to the unscaled potential (\ref{eq:potexp.utr}).
That unscaled potential is  independent of $\Lambda$, as it must be since $K$
does not depend on $\Lambda$.

Of more interest for nuclear physics are nontrivial fixed points that can
describe systems with very strong scattering at low energies. The simplest of
these can be found by considering a potential that depends only on energy,
$\hat V=\hat V_0(\hat p)$. This will be a fixed-point solution to the full RG
equation (\ref{eq:scrge}) if it satisfies
\begin{equation}\label{eq:fprge.ere}
\hat p{\partial\hat V_0\over\partial\hat p}
+\hat V_0(\hat p)+{\hat V_0(\hat p)^2\over 1-\hat p^2}=0.
\end{equation}
Solving this equation subject to the boundary condition that the potential be 
analytic in $\hat p^2$ as $\hat p^2\rightarrow 0$ we obtain
\begin{equation}
\hat V_0(\hat p)=-\left[1-{\hat p\over 2}\ln{1+\hat p\over 1-\hat p}
\right]^{-1}.
\end{equation}
Note that, although the detailed form of the energy-dependence of this
potential is determined by our particular choice of cut-off, the fact that it
tends to a constant as $\hat p\rightarrow 0$ is a generic feature. The
corresponding unscaled potential is
\begin{equation}\label{eq:potfp.ere}
V_0(p,\Lambda)=-{2\pi^2\over M}\left[\Lambda-{p\over 2}
\ln{\Lambda+p\over\Lambda-p}\right]^{-1}.
\end{equation}
At $p=0$, this potential is inversely proportional to $\Lambda$, a property
which holds for any form of cut-off.

For a momentum-independent potential like $V_0$ the LS equation takes a
particularly simple form. For $V_0$ we find that its solution for $K$ is
infinite, or rather $1/K=0$. This corresponds to a system with infinite
scattering length, or equivalently a bound state at exactly zero energy.

To study the behaviour near this fixed point we consider small perturbations
about it that scale with definite powers of $\Lambda$: 
\begin{equation}
\hat V(\hat k',\hat k,\hat p,\Lambda)=\hat V_0(\hat p)
+C\Lambda^\nu \phi(\hat k',\hat k,\hat p).
\end{equation}
These satisfy the linearised RG equation
\begin{eqnarray}\label{eq:linrge.ere}
\hat k'{\partial\phi\over\partial\hat k'}
&+&\hat k{\partial\phi\over\partial\hat k}
+\hat p{\partial\phi\over\partial\hat p}+\phi\\
&+&{\hat V_0(\hat p)\over 1-\hat p^2}\left[
\phi(\hat k',1,\hat p)+\phi(1,\hat k,\hat p)\right]
=\nu\phi.\nonumber
\end{eqnarray}

Let us first look for solutions to Eq.~(\ref{eq:linrge.ere}) that depend only
on energy ($p^2$). The equation can be integrated straightforwardly, making use
of the fixed point equation for $\hat V_0$ (\ref{eq:fprge.ere}). The solutions
are
\begin{equation}\label{enper.ere}
\phi(\hat p)=\hat p^{\nu+1} \hat V_0(\hat p)^2.
\end{equation}
If we demand that these be well-behaved functions of $p^2$ as $p^2\rightarrow 
0$, then we find that the allowed RG eigenvalues are $\nu=-1,1,3,\dots$. From
this we see that the fixed point is unstable: it has one negative eigenvalue.

The RG flow corresponding to Eq.~(\ref{eq:scrge}) is shown in Fig.~1, projected 
into the plane corresponding to the first two terms in the expansion of the 
potential in powers of energy, $\hat V(\hat p)=b_0+b_2\hat p^2+\cdots$. This 
shows the two fixed points discussed here: the trivial one at the origin and
the nontrivial one at $b_0=b_2=-1$. The precise position of the latter point
depends on our particular choice of cut-off, but the pattern of the RG flow is 
general. Potentials that lie exactly on the ``critical surface" ($b_0=-1$ 
for our sharp cut-off) flow towards the nontrivial fixed point as 
$\Lambda\rightarrow 0$. For a system with a small perturbation away from this 
surface, the potential initially flows towards the nontrivial fixed point
for large $\Lambda$. Eventually, however, the unstable perturbation 
causes these potentials to flow either to the trivial fixed point or to 
infinity.

\begin{figure}[h,t,b,p]
\vspace{0.5cm} \epsfysize=8cm \centerline{\epsffile{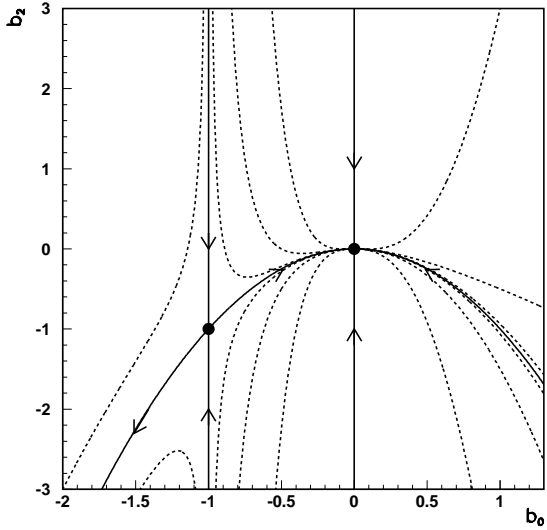}}
\centerline{{\caption{\label{fig:rgflow} The RG flow
of the first two terms in the expansion of the rescaled potential in powers of 
energy. The two fixed points are indicated by the black dots. The solid lines
are flow lines that approach one of the fixed points along a direction 
corresponding to an RG eigenfunction; the dashed lines are more general
flow lines. The arrows indicate the direction of flow as $\Lambda\rightarrow
0$. }}}
\end{figure}

The unscaled potential including the perturbations (\ref{enper.ere}) is
\begin{eqnarray}\label{eq:pot.ere}
&&V(k',k,p,\Lambda)\\
&&\quad=V_0(p,\Lambda)
+{M\Lambda_0\over 2\pi^2}\sum_{n=0}^\infty\widehat C_{2n-1}
\left({p\over\Lambda_0}\right)^{2n}V_0(p,\Lambda)^2.\nonumber
\end{eqnarray}
The (on-shell) $K$-matrix for this potential has the effective-range expansion
\begin{equation}\label{eq:enere.ere}
\frac{1}{K(p,p,p)}=-{M\Lambda_0\over 2\pi^2}
\sum_{n=0}^\infty\widehat C_{2n-1}\left({p\over\Lambda_0}\right)^{2n}+\cdots.
\end{equation}
To first order in the coefficients $\widehat C_\nu$ of the eigenfunctions,
we see that the terms in the expansion of the potential are in one-to-one 
correspondence with the terms in the effective-range expansion. In particular,
$\widehat C_{-1}$ and $\widehat C_1$ are given in terms of the scattering
length and effective range by
\begin{equation}
\widehat C_{-1}=-{\pi\over 2\Lambda_0 a},\qquad 
\widehat C_1={\pi\Lambda_0 r_e\over 4}.
\end{equation}

At this point we can compare our potential (\ref{eq:pot.ere}) with that found
by KSW in the PDS scheme\cite{ksw2}. To first order in $1/a$ and $p^2$ their
potential can be written in the form
\begin{equation}\label{eq:potksw}
V(p,\mu)={4\pi\over M\mu}\left[-1-{1\over\mu a}-{r_e\over 2\mu}p^2
+\cdots\right].
\end{equation}
Remembering that the scale $\mu$ in a subtractive renormalisation scheme acts 
like a resolution scale and so plays an analogous role to the cut-off 
$\Lambda$, we see that the $1/\mu$ dependence of the first term in 
Eq.~(\ref{eq:potksw}) agrees with the $1/\Lambda$ dependence of the 
fixed-point potential (\ref{eq:potfp.ere}). Similarly 
the factors of $1/\mu^2$ in the second and third terms agree with the 
$1/\Lambda^2$ factors in the energy-dependent perturbations in 
Eq.~(\ref{eq:pot.ere}). If, as for perturbations around the trivial fixed 
point, we assign an order $d=\nu-1$ to each term in the potential, then we see
that the power counting for (energy-dependent) perturbations around the 
nontrivial fixed point agrees with that of KSW.

For systems with finite but large scattering lengths, it is still possible to
build a description based on Weinberg's expansion around the trivial fixed
point. To do so one must solve the RG equation to all order in the scattering
length, resumming terms involving powers of $\Lambda a$. This is approach
adopted by van Kolck\cite{vk2,vk3}. It can be pictured as following one the
flow lines in Fig.~1 that approach the trivial fixed point close to the line
that connects the two fixed points. For values of $\Lambda$ that are large
compared with $1/a$ such a flow line lies close to the critical surface for the 
nontrivial fixed point and so the behaviour of the system can be organised
according to the power counting associated with that fixed point.

So far we have considered perturbations of the potential that depend only on
energy, but there are also ones that depend on momentum as well. To find
these, we look for solutions (\ref{eq:fprge.ere}) that have the form 
\begin{equation}
\phi(\hat k',\hat k,\hat p)=\hat k^n\phi_1(\hat p)+\phi_2(\hat p). 
\end{equation}
(A Hermitian potential can be obtained by adding a similar term with $k
\rightarrow k'$.) In this case the solutions are
\begin{equation}\label{eq:mompert.ere}
\phi(\hat k',\hat k,\hat p)=
\left[\hat k^n-\hat p^n+\sum_{m=0}^{n/2-1}{\hat p^m\over n-2m+1}
\hat V_0(\hat p)\right]\hat V_0(\hat p),
\end{equation}
with RG eigenvalues $\nu=n=2,4,6\dots$. Multiplying any of these functions
by $p^m$ where $m$ is a positive even integer also gives an eigenfunction,
with $\nu=n+m$. An important point to note is that the momentum-dependent 
eigenfunctions have different eigenvalues from the corresponding purely 
energy-dependent ones. This is quite unlike the more familiar case of
perturbations around the trivial fixed point where, for example, the $p^2$ and
$k^2$ terms in the potential are both of the same order, $\nu=3$. It means
that, in the vicinity of the nontrivial fixed point, one cannot make a field
transformation to eliminate energy dependence from the potential in
favour of momentum dependence without introducing a much more complicated
cut-off dependence into the the effective potential.

To complete the picture, we note that it is possible to find solutions to
Eq.~(\ref{eq:linrge.ere}) that depend on both $k$ and $k'$. These are products
of two factors of the form of the expression in square brackets in
Eq.~(\ref{eq:mompert.ere}), one depending on $k$ and one on $k'$, and they
have RG eigenvalues $\nu=n+n'+1=5,7,9\dots$. Each of these can also be
multiplied by even powers of $p$ to yield further eigenfunctions. Together,
the eigenfunctions described above contain all possible products of powers of
$k^2$, $k^{\prime 2}$ and $p^2$. They thus form a complete set that can be used
to expand any perturbation about the fixed-point potential that is well-behaved 
as $k^2,k^{\prime 2},p^2\rightarrow 0$ and so has a power-series expansion in
these quantities.

When the $\nu=2$ momentum-dependent perturbation is included, the term
\begin{equation}
\widehat C_2\left\{\left[k^2-p^2+{1\over 3}{M\Lambda^3\over 2\pi^2}
V_0\right]
{V_0(p,\Lambda)\over\Lambda_0^2}+\;(k\rightarrow k')\right\}
\end{equation} 
must be added to the potential (\ref{eq:pot.ere}). The resulting potential has 
a two-term separable structure and so the LS equation can be solved using the 
techniques in Refs.\cite{md2,rbm}. To first order in the $\widehat C_\nu$, we
find that the effective-range expansion is again given by
Eq.~(\ref{eq:enere.ere}); the momentum-dependent perturbation does not
contribute to the on-shell scattering. This can be understood from the fact 
that its coefficient involves an even power of the underlying scale 
$\Lambda_0$, whereas all of the terms in the effective range expansion 
contain odd powers of that scale.

The identification of the terms in the potential and the effective-range 
expansion is straightforward at first order in the coefficients $\widehat
C_\nu$ because only energy dependent perturbations contribute to the
scattering. To check whether this equivalence persists to higher order in the
$\widehat C_\nu$, we have solved the RG equation to second order in these
coefficients. To illustrate the behaviour, we consider here corrections to the
potential (\ref{eq:pot.ere}) up to order $\widehat C_{-1}\widehat C_{2}$. We
find the following additional pieces
\begin{equation}\label{eq:pot2.ere}
\frac{M}{2\pi^2}\widehat C_{-1}\widehat C_{2}\left(
k'^2+k^2+A\;p^2+\frac{4}{3}\frac{M\Lambda^3}{2\pi^2}V_0\right)
\frac{V_0(p,\Lambda)^2}{\Lambda_0},
\end{equation}
where $A$ is a constant of integration which is not fixed by the boundary
conditions. This undetermined piece arises from the solution of the
homogeneous part of the linearised RG equations, and has the exactly same
structure as the $\nu=1$ term in (\ref{eq:pot.ere}).  

This second-order piece (\ref{eq:pot2.ere}) will in general contribute to the 
effective range, along with the $\nu=1$ term. The direct correspondence between 
scattering observables and terms in the potential can be maintained by choosing 
$A=-2$, which ensures that the contribution of $\widehat C_{-1}\widehat C_{2}$ 
to the effective range vanishes.

Alternatively, one could set $\widehat C_1=0$ (and $A=0$) and generate the 
effective range entirely from the $\nu=2$ momentum-dependent perturbation. 
However the required coefficient in this case is
\begin{equation}
\widehat C_2={\Lambda_0^2ar_e\over 4},
\end{equation}
implying the existence of large factors of $\Lambda a$ that need to be 
resummed in the potential. This can be done by starting from the trivial fixed
point, where energy dependence can be eliminated in favour of momentum
dependence, and then using the approach of van Kolck\cite{vk2,vk3} to follow
the RG flow back to the vicinity of the critical surface for the nontrivial
fixed point. However the resummation procedure tends to mask the simple nature 
of the RG flow close to the nontrivial fixed point.

We have seen that, to second order in the coefficients $\widehat C_\nu$ 
in the expansion about the nontrivial fixed point, only the energy-dependent 
RG eigenfunctions contribute to the on-shell scattering. So long as analogous 
procedures can be carried out to all orders, the effective theory defined by an 
expansion around the nontrivial fixed point is systematic and the terms in the 
potential are in one-to-one correspondence with those of the effective-range 
expansion. The  resulting potential is determined entirely by on-shell 
scattering observables. Indeed, as has long been known from the effective-range 
expansion, the scattering length and effective range are also sufficient to 
determine the asymptotic $s$-wave part of the deuteron wave function\cite{bl}. 

Our RG analysis thus demonstrates that there is an equivalence between
the effective field theory based on the nontrivial fixed point and the
effective-range expansion \cite{ere}, as previously suggested by van
Kolck\cite{vk1,vk3}.

The effective-range expansion is based on
\begin{equation}
p\cot\delta(p)=-{4\pi\over M}{1\over K(p,p,p)},
\end{equation}
which is the logarithmic derivative at the origin of the asymptotic wave 
function. It can thus be regarded as an energy-dependent boundary condition on
the wave function at the origin. As noted by van  Kolck\cite{vk1,vk3}, this
corresponds to the fact that the parts of the  effective field theory which
contribute to observables act like a quasipotential\cite{bre}. 

The fixed point is, as already noted, an unstable one. Only potentials that
lie on the ``critical surface" defined by $1/a=0$ (corresponding to a bound
state at zero energy) will flow to the fixed point as $\Lambda\rightarrow 0$.
For small but nonzero values of $1/a$ the potential will eventually either 
tend to the trivial fixed point or diverge to infinity, depending on the sign
of $a$. Nonetheless for cut-offs in the range $1/a\ll\Lambda\ll\Lambda_0$ the
behaviour of the potential may be dominated by the flow towards the nontrivial
fixed point. In such a regime we can still use the eigenfunctions found above
to define a systematic expansion of the potential, as noted in Ref.\cite{ksw2}.

To summarise: by applying the renormalisation group to two-body scattering
we have identified two fixed points. One is the trivial fixed point describing 
perturbative scattering. The other describes systems with bound states at
zero energy and is directly related to the effective-range
expansion\cite{ere}. By studying the eigenfunctions of the linearised RG we
can assign a systematic power-counting to terms in the expansion of the
potential around each of these fixed points. In the case of the nontrivial
fixed point, each term in this expansion corresponds to a term in the
effective-range expansion. The potential to next-to-leading order depends only
on energy and is determined entirely by on-shell scattering observables, namely
the scattering length and effective range. The success of the effective-range
expansion\cite{ere,bl} can be therefore be understood in terms of an effective
field theory based on this nontrivial fixed point\cite{bvk,pkmr,ksw2}. It also 
suggests that there should be ways to extend this expansion to treat 
long-ranged forces such as pion-exchange\cite{ksw2}, as well as three-body
systems\cite{bvk}, both of which are being actively pursued.

We are grateful to D. Phillips and U. van Kolck for helpful discussions. MCB
thanks the organisers of the Caltech/INT workshop which provoked this line of
investigation. This work was supported by the EPSRC and PPARC.

\end{document}